\documentclass[aps,prl,superscriptaddress,nofootinbib,floatfix,twocolumn,longbibliography]{revtex4-1}
\RequirePackage{lineno}
\usepackage[colorlinks,linkcolor=blue,citecolor=blue,filecolor=black,urlcolor=blue]{hyperref}
\usepackage{epsfig,graphics}
\usepackage{graphicx}
\usepackage{bm}
\usepackage{amssymb}
\usepackage{amsmath}
\usepackage{multirow}
\usepackage{MnSymbol}
\usepackage{ragged2e}
\usepackage{moresize}
\makeatletter 
\@addtoreset{equation}{section}
\makeatother  

\usepackage{caption}
\DeclareCaptionLabelFormat{nature}{\textbf{#1\ #2}\ $\vert$\ }
\newcommand{\lr}[1]{\left\langle #1\right\rangle}
\newcommand{\llrr}[1]{\left\llangle #1\right\rrangle}
\newcommand{\pT} {\ensuremath{p_{\mathrm{T}}}}
\newcommand{\nchrec}{\mbox{$N_{\mathrm{ch}}^{\mathrm{rec}}$}}
\newcommand{\nch}{N_{\mathrm{ch}}}
\newcommand{\snn}{\mbox{$\sqrt{s_{\mathrm{NN}}}$}}
\begin{document}
\title{Imaging Shapes of Atomic Nuclei in High-Energy Nuclear Collisions}
\author{STAR Collaboration}
\begin{abstract}
Atomic nuclei are self-organized, many-body quantum systems bound by strong nuclear forces within femtometer-scale space. These complex systems manifest a variety of shapes~\cite{bohr,Heyde:2016sop}, traditionally explored using non-invasive spectroscopic techniques at low energies~\cite{Cline:1986ik,Yang:2022wbl}. However, at these energies, their instantaneous shapes are obscured by long-timescale quantum fluctuations, making direct observation challenging. Here we introduce the ``collective flow assisted nuclear shape imaging'' method, which images the nuclear global shape by colliding them at ultrarelativistic speeds and analyzing the collective response of outgoing debris. This technique captures a collision-specific snapshot of the spatial matter distribution within the nuclei, which, through the hydrodynamic expansion, imprints patterns on the particle momentum distribution observed in detectors~\cite{Shuryak:2014zxa,Busza:2018rrf}. We benchmark this method in collisions of ground state Uranium-238 nuclei, known for their elongated, axial-symmetric shape. Our findings show a large deformation with a slight deviation from axial symmetry in the nuclear ground state, aligning broadly with previous low-energy experiments. This approach offers a new method for imaging nuclear shapes, enhances our understanding of the initial conditions in high-energy collisions and addresses the important issue of nuclear structure evolution across energy scales.
\end{abstract}
\maketitle

More than 99.9\% of the visible matter in the cosmos resides in the center of atoms -- the atomic nuclei composed of nucleons (protons and neutrons). Our knowledge of their global structure primarily comes from spectroscopic or scattering experiments~\cite{Cline:1986ik,Yang:2022wbl,Hofstadter:1956qs} at beam energies below hundreds of MeV per nucleon. These studies show that most nuclei are ellipsoidally deformed, with greater deformation in nuclei distant from magic numbers (2, 8, 20, 28, 50, 82, and 126)~\cite{Moller:2015fba}. Investigating nuclear shape across the Segr\`e chart has been an important area of research over many decades and is crucial for topics such as nucleosynthesis~\cite{Schatz:1998zz}, nuclear fission~\cite{Schunck:2022gwo}, and neutrinoless double beta decay ($0\nu\beta\beta$)~\cite{Engel:2016xgb}.

\begin{figure*}[htbp]
\centering
\includegraphics[width=1\linewidth]{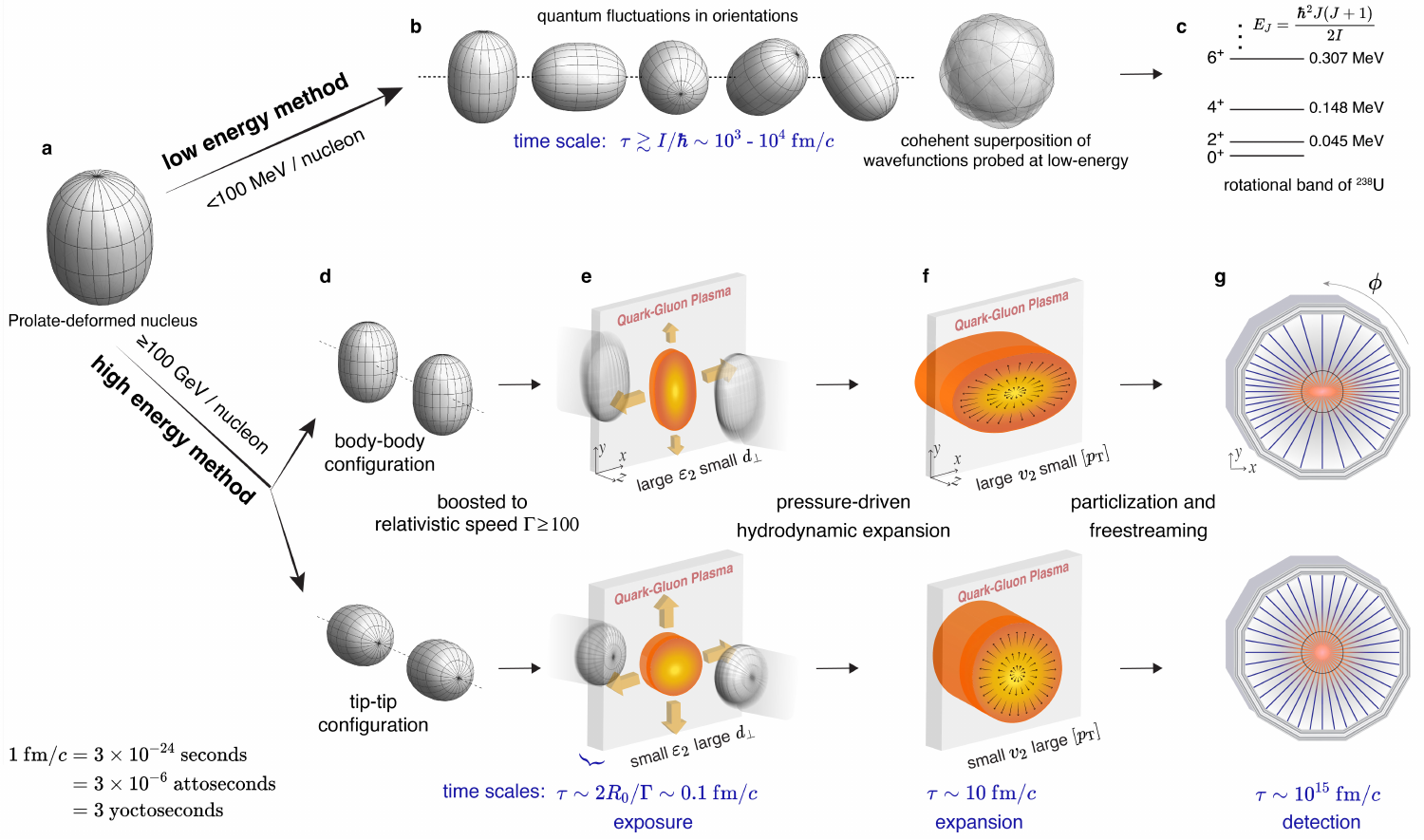}
\caption{\label{fig:1}\textbf{Methods for determining the nuclear shape in low and high energies}. \textbf{a,} Cartoon of a well-deformed prolate-shaped nucleus. \textbf{b,} Quantum fluctuations over Euler angles for this nucleus and associated overall timescale. \textbf{c,} Quantum mechanical manifestation of the deformation in terms of the first rotational band of $^{238}$U. \textbf{d,} Aligning the two nuclei in head-on body-body configuration (top) and tip-tip configuration (bottom). \textbf{e,} High-energy collision of two Lorentz-contracted nuclei and resulting 3D profile of the initially-produced quark-gluon plasma, in which the arrows indicate the pressure gradients. \textbf{f,} The 3D profile of the quark-gluon plasma at the end of the hydrodynamic expansion before it freezes out into particles, in which arrows indicate the velocities of fluid cells.  \textbf{g,} Charged particle tracks measured in the detector. The timescales shown are in units of fm/$c$ -- the time for light to travel 1 femtometer. The body-body configuration has large eccentricity $\varepsilon_2$ and small gradient $d_{\perp}$, leading to large elliptic flow $v_2$ and smaller average transverse momentum $[\pT]$, and vice versa for tip-tip configuration (see main text).}
\vspace*{-0.5cm}
\end{figure*}

In a collective model picture, the ellipsoidal shape of a nucleus with mass number $A$ is defined in the intrinsic (body-fixed) frame, in which its surface $R(\theta,\phi)$ is described by~\cite{bohr}, 
\begin{align}\label{eq:1}
R(\theta,\phi) &= R_0(1+\beta_2 [\cos \gamma Y_{2,0}+ \sin\gamma Y_{2,2}])\;.
\end{align}
Here $R_0\approx 1.2A^{1/3}$ fm represents the nuclear radius. The spherical harmonics in the real basis $Y_{l,m}(\theta,\phi)$, the quadrupole deformation magnitude $\beta_2$, and the triaxiality parameter $\gamma$ define the nuclear shape.  The $\gamma$ parameter, spanning $0^{\circ}$--$60^{\circ}$, controls the ratios of principal radii. Specifically, $\gamma=0^{\circ}$ corresponds to a prolate shape, $\gamma=60^{\circ}$ an oblate shape, and values in between $0^{\circ}<\gamma<60^{\circ}$ to a triaxial shape.  Although most nuclei are axially-symmetric (prolate or oblate) or have a fluctuating $\gamma$ value ($\gamma$-soft), the rigid triaxial shape is uncommon~\cite{Toh:2013qba}. An example of an axial-symmetric, prolate-deformed nucleus is shown in Fig.~\ref{fig:1}a.

Nuclear shapes, even in ground states, are not fixed. They exhibit zero-point quantum fluctuations involving various collective and nucleonic degrees of freedom (DOF) at different timescales. These fluctuations superimpose on each other in the laboratory frame. In well-deformed nuclei like $^{238}$U, dominant fluctuations are in the rotational DOF with a timescale of $\tau_{\mathrm{rot}}\sim I/\hbar \sim 10^{3}$--$10^{4}$ fm/$c$ (1~fm/$c=3\times 10^{-24}$ seconds~$=$~3 yoctoseconds)~\cite{Nakatsukasa:2016nyc}, where $I$ denotes the moment of inertia (Fig.~\ref{fig:1}b). Consequently, measurement processes in spectroscopic methods, lasting orders of magnitude longer than $\tau_{\mathrm{rot}}$, capture a coherent superposition of wavefunctions in all orientations. Their shapes are usually inferred by comparing spectroscopic data (Fig.~\ref{fig:1}c) with model calculations~\cite{RevModPhys.83.1467,Bender:2003jk}. Traditional electron-nucleus scattering experiments, although faster than $\tau_{\mathrm{rot}}$, probe mainly localized regions of the nucleus, giving an orientation-averaged spherical image after accumulating many events, in which the impact of deformation manifests as a broadening of the charge distribution~\cite{Hofstadter:1956qs,bohr}.

\textbf{New shape-imaging method}. To directly observe the global shape of the nuclei, a measurement must 1) be much quicker than $\tau_{\mathrm{rot}}$ and 2) provide access to the many-body nucleon distribution in each nucleus. High-energy nuclear collisions, an utterly destructive process, remarkably fulfil these criteria. Conducted at the Relativistic Heavy-Ion Collider (RHIC) and the Large Hadron Collider (LHC) with center-of-mass energies per nucleon pair ($\snn$) reaching up to 5000 GeV, these collisions completely obliterate the nuclei, temporarily forming a quark-gluon plasma (QGP) -- a hot, dense matter of interacting quarks and gluons~\cite{Shuryak:2014zxa,Busza:2018rrf}. The nuclear shape influences the geometry of QGP and its collective expansion, imprinting itself on the momentum distribution of the produced particles. In an ironic twist, this effectively realizes Richard Feynman's analogy of the seemingly impossible task of ``figuring out a pocket watch by smashing two together and observing the flying debris.'' The collective response plays a key role.

Our shape-imaging technique focuses on head-on (near zero impact parameter) collisions of prolate-deformed nuclei (Fig.~\ref{fig:1}d-g). The initial configurations lie between two extremes: body-body (top) and tip-tip (bottom) collisions (Fig.~\ref{fig:1}d). Before impact, Lorentz-contraction flattens the ground-state nuclei into pancake-like shapes by a factor of $\Gamma= \frac{1}{2}\snn/m_{0}> 100$, where $m_0\approx 0.94$ GeV is the nucleon mass (Fig.~\ref{fig:1}e). The initial impact, lasting $\tau_{\mathrm{expo}}=2R_0/\Gamma\lesssim 0.1$~fm/$c$, acts as an exposure time. The shape and size of the overlap region, reflecting the initially produced QGP, directly mirror those of the colliding nuclei projected on the transverse ($xy$) plane (Fig.~\ref{fig:1}e). Body-body collisions create a larger, elongated QGP, which undergoes pressure-gradient-driven expansion (indicated by arrows) until about 10 fm/$c$~\cite{Busza:2018rrf}, resulting in an inverted, asymmetric distribution (Fig.~\ref{fig:1}f). By contrast, tip-tip collisions form a compact, circular QGP, driving a more rapid but symmetric expansion (Fig.~\ref{fig:1}f). In the final stage, the QGP freezes into thousands of particles, captured as tracks in detectors, whose angular distributions reflect the initial QGP shape (Fig.~\ref{fig:1}g). This flow-assisted imaging is similar to the Coulomb explosion imaging in molecule structure analysis~\cite{cei,cei2,cei3,ceinature,ceinatureb}, in which the spatial arrangement of atoms, ionized by an X-ray laser or through passage in thin foils, is deduced from their mutual Coulomb-repulsion-driven expansion. However, the expansion duration in high-energy collisions is $10^{6}$--$10^{9}$ times shorter.

The concept that the dynamics of QGP can be utilized to image the geometrical properties of its initial condition was widely recognized. This understanding has facilitated the determination of the collision impact parameter and fluctuations in nucleon positions~\cite{Busza:2018rrf}, as well as the neutron skin of the colliding nuclei~\cite{Giacalone:2023cet}, by measuring higher-order harmonic flows.  However, our study took a further step to image the shape of the colliding nuclei through their impact on the initial condition.

\textbf{Energy evolution of shapes?} A pertinent question is how the shapes observed in high-energy colliders compare with those derived from low-energy experiments. For well-deformed nuclei like $^{238}$U, we expect them to align at a basic level. However, there are other correlations (such as clustering and short-range correlations) that manifest at increasingly faster timescales from 1000 to a few fm/$c$. Moreover, high-energy collisions also probe nuclear structure at sub-nucleonic levels, such as quark and gluon correlations, as well as modifications caused by dense gluon fields~\cite{Mantysaari:2023qsq}. As a result, the deformations observed at high energy may differ from those at low energy, motivating us to examine nuclear phenomena across energy scales and discover new phenomena.

\textbf{Observables}. In Fig.~\ref{fig:1}e, the QGP's initial shape is quantified by the eccentricity, $\varepsilon_2=\frac{\lr{y^2}-\lr{x^2}}{\lr{y^2}+\lr{x^2}}$, calculated from the nucleon distribution in the $xy$-plane, perpendicular to the beam direction. The hydrodynamic expansion, reacting to $\varepsilon_2$, results in particle anisotropy, described as \mbox{$dN/d\phi~\propto~1+2v_2\cos(2\phi)$} aligned with the impact parameter along the $x$-axis.  This phenomenon, known as ``elliptic flow'' ($v_2$)~\cite{Ollitrault:1992bk}, is shown in Fig.~\ref{fig:1}g. Moreover, the compactness of the QGP, indicated by the inverse area of the overlap $d_{\perp} \propto 1/\sqrt{\lr{x^2}\lr{y^2}}$~\cite{Schenke:2020uqq}, influences the radial expansion or ``radial flow'', captured in the event-wise average transverse momentum ($[\pT]$). A key discovery at RHIC was the behavior of QGP as a nearly perfect, inviscid fluid~\cite{STAR:2005gfr,PHENIX:2004vcz}, effectively transforming initial geometry into final state anisotropies. Hydrodynamic models have confirmed linear response relations: $v_2\propto \varepsilon_2$~\cite{Niemi:2012aj} and $\delta \pT \propto \delta d_{\perp}$~\cite{Bozek:2012fw}, where $\delta \pT = [\pT]-\lr{[\pT]}$ and $\delta d_{\perp} = d_{\perp}-\lr{d_{\perp}}$ denote event-wise deviations from mean values.

In head-on collisions with spherical nuclei, non-zero $\varepsilon_2$ and $\delta d_{\perp}$ can be generated by the random fluctuations in the position of nucleons in the overlap region. In non-head-on collisions, besides these stochastic elements, the overlap region also has an average elliptical shape. This average shape significantly contributes to $\varepsilon_2$, known as reaction plane eccentricity $\varepsilon_2^{\mathrm{rp}}$~\cite{PHOBOS:2007vdf} but has little effect on the radial quantity $\delta d_{\perp}$.

Prolate deformation further modifies $\varepsilon_2$ and $d_{\perp}$. Body-body collisions in this context yield high $\varepsilon_2$ and low $d_{\perp}$ values, and vice versa for tip-tip collisions. This leads to enhanced, anti-correlated event-by-event fluctuations in $\varepsilon_2$ and $d_{\perp}$~\cite{Giacalone:2019pca}, measurable through observables such as $\lr{v_2^2}$, $\lr{(\delta \pT)^2}$, and $\lr{v_2^2\delta \pT}$~\cite{Bozek:2016yoj} that are linearly related to the moments of the initial condition $\lr{\varepsilon_2^2}$, $\lr{(\delta d_{\perp})^2}$, and $\lr{\varepsilon_2^2\delta d_{\perp}}$. These observables, linked to two- and three-body nucleon distributions in the intrinsic frame (Methods), were found to have a simple parametric dependence on shape parameters~\cite{Jia:2021qyu}:
\begin{align}\nonumber
\lr{v_2^2}&= a_1+b_1\beta_2^2\;,\\\nonumber
\lr{(\delta \pT)^2}&= a_2+b_2\beta_2^2\;,\\\label{eq:2}
\lr{v_2^2\delta \pT} &= a_3-b_3\beta_2^3\cos(3\gamma)\;.
\end{align}
The positive coefficients $a_n$ and $b_n$ capture the collision geometry and QGP properties. The $b_n$ values are nearly independent of the impact parameter, while $a_n$ values are minimized in head-on collisions, making such collisions ideal for constraining nuclear shape. Our study offers the first quantitative and simultaneous determination of $\beta_{2}$ and $\gamma$ using all three observables in Eq.~\eqref{eq:2}. 

\textbf{Nuclear shapes from low energy data}.  Our measurements use data from high-energy $^{238}$U+$^{238}$U and $^{197}$Au+$^{197}$Au collisions. These species have contrasting shapes: mildly oblate $^{197}$Au (close to magic numbers with $Z$=79 protons and $N$=118 neutrons) and highly prolate $^{238}$U (an open shell nucleus with 92 protons and 146 neutrons). This comparison helps us to deduce the shape of $^{238}$U. A state-of-the-art beyond mean-field model, which reproduces the bulk of experimental data on $^{197}$Au, predicts deformation values of $\beta_{2\mathrm{Au}}\approx 0.12$--0.14 and $\gamma_{\mathrm{Au}}\approx 43^{\circ}$~\cite{Bally:2023dxi}. The deformation of $^{238}$U, inferred from measured transition rates within rotational spectra, is estimated at $\beta_{2\mathrm{U}}=0.287\pm0.007$~\cite{Pritychenko:2013gwa}. 

Experimental estimates of the Uranium triaxiality have been derived from energy levels and transition data under a rigid-rotor assumption, suggesting $\gamma_{\mathrm{U}}=6^{\circ}$--$8^{\circ}$~\cite{PhysRevC.54.2356}. One important issue concerns the softness of $\gamma$: whether the nuclei have rigid triaxial shape or fluctuations of $\gamma$ around its mean value~\cite{Sharpey-Schafer:2019ajz}. This issue is complicated by possible changes of $\gamma$ when nuclei are excited~\cite{Otsuka:2023yts}. Our three-body observable $\lr{v_2^2\delta \pT}$ in Eq.~\eqref{eq:2} is sensitive only to the mean of the triaxiality, not its fluctuations~\cite{Dimri:2023wup}. Nevertheless, measuring $\beta_{2\mathrm{U}}$ and $\gamma_{\mathrm{U}}$ could validate our imaging method and investigate its ground-state triaxiality.

\begin{figure*}[htbp]
\centering
\includegraphics[width=0.8\linewidth]{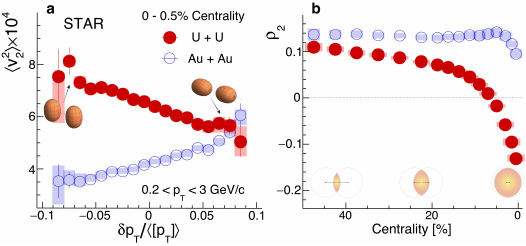}
\caption{\label{fig:2} \textbf{Correlation between elliptic flow and radial flow}. \textbf{a,} $\lr{v_2^2}$ versus $\delta \pT/\lr{[\pT]}$ in 0--0.5\% most central Au+Au and U+U collisions. \textbf{b,} $\rho_2~=~\frac{\lr{v_2^2 \delta \pT}}{\lr{v_2^2}\sqrt{\lr{(\delta \pT)^2}}}$ across centrality, quantifying the strength of $v_2$--$\delta \pT$ correlation. The elliptic-shaped overlaps in the transverse plane for various centralities are also shown.}
\vspace*{-0.5cm}
\end{figure*}

\textbf{Experimental setup}. Our analysis uses U+U data from 2012 and Au+Au data from 2010 and 2011 at $\snn= 193$ GeV and 200 GeV, respectively, using the STAR detector at RHIC. Each collision produces up to 2000 charged particles in the STAR time-projection chamber (TPC)~\cite{Anderson:2003ur}, covering the polar angle range $|\theta-90^{\circ}|\lesssim50^{\circ}$ and full $\phi$ range. The TPC tracks these particles and determines their $\pT$. Collision events are categorized by ``centrality'', defined as the percentage of the total inelastic cross-section, with lower percentages indicating a larger number of created particles. We calculate $\lr{v_2^2}$, $\lr{(\delta\pT)^2}$, and $\lr{v_2^2\delta\pT}$ by established methods~\cite{Bilandzic:2010jr} using tracks in $0.2<\pT<3$ GeV/$c$. The results incorporate the uncertainties arising from track selection, reconstruction efficiency, background events, and correlations unrelated to flow (Methods).

\textbf{Results}. To directly observe the shape-size correlation depicted in Fig.~\ref{fig:1}e--f, we analyze the 0--0.5\% most central collisions and correlate $\lr{v_2^2}$ with event-wise $\delta \pT$ values (Fig.~\ref{fig:2}a). A pronounced anticorrelation in U+U collisions aligns with the expectation~\cite{Giacalone:2019pca}: events with small (large) $\delta\pT$ are enriched with body-body (tip-tip) collisions. This effect is striking, as $\lr{v_2^2}$ in U+U is twice that of Au+Au at the lowest $\delta \pT$, yet similar at the highest $\delta \pT$. 

We quantify this correlation using normalized covariance $\rho_2~=\frac{\lr{v_2^2 \delta \pT}}{\lr{v_2^2}\sqrt{\lr{(\delta \pT)^2}}}$ (Fig.~\ref{fig:2}b). In Au+Au collisions, $\rho_2$ is relatively constant, with a minor decrease in the central region due to centrality smearing~\cite{ATLAS:2022dov}. This smearing can be reduced by averaging over a wider, say 0--5\% centrality range~\cite{Zhou:2018fxx}. By contrast, $\rho_{2}$ in U+U collisions decreases steadily, turning negative at about 7\% centrality, reflecting the large prolate deformation of $^{238}$U. The deformation has the greatest impact in central collisions but also influences other centrality ranges.

\begin{figure*}[htbp]
\includegraphics[width=1\linewidth]{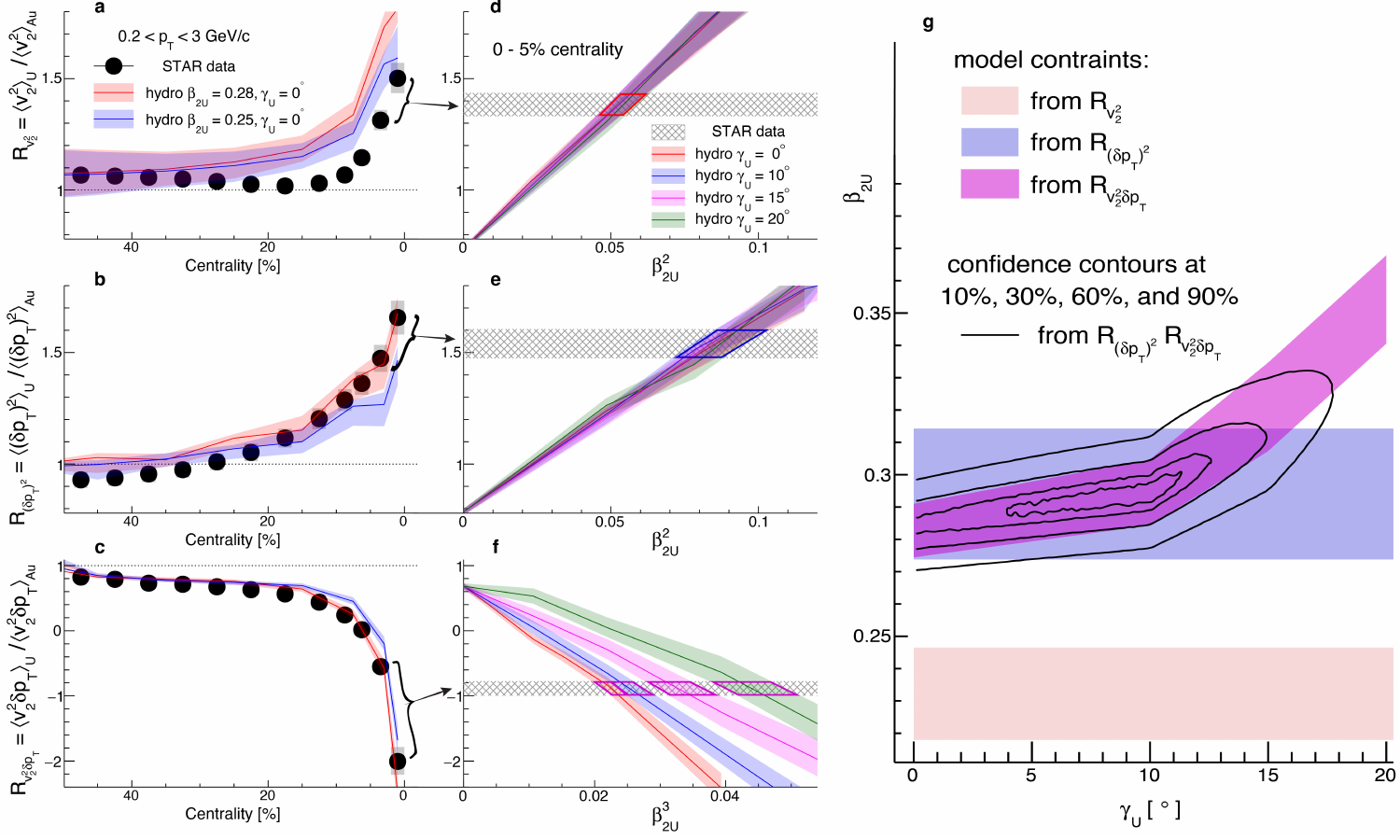}
\caption{\label{fig:3} 
\textbf{Constraining the shape of $^{238}$U}. Left column: ratios of $\lr{v_2^2}$ (\textbf{a}), $\lr{(\delta\pT)^2}$ (\textbf{b}), and $\lr{v_2^2\delta\pT}$ (\textbf{c}) between U+U and Au+Au collisions as a function of centrality. The data are compared with the IP-Glasma+MUSIC hydrodynamic model calculation assuming $\beta_{2\rm U}=0.28$ (red) and $\beta_{2\rm U}=0.25$ (blue), shaded bands of which denote the model uncertainties (Methods). Mid-column: ratio values in 0--5\% most central collisions (hatch bands) for $\lr{v_2^2}$ (\textbf{d}), $\lr{(\delta\pT)^2}$ (\textbf{e}), and $\lr{v_2^2\delta\pT}$ (\textbf{f}) are compared with model calculations as a function of $\beta_{2\mathrm{U}}^2$ or $\beta_{2\mathrm{U}}^3$ for four $\gamma_{\mathrm{U}}$ values. The colored quadrilaterals delineate the allowed ranges of $\beta_{2\mathrm{U}}^2$ or $\beta_{2\mathrm{U}}^3$ from this data-model comparison. \textbf{g}, the constrained ranges of ($\beta_{2\mathrm{U}}$,$\gamma_{\mathrm{U}}$) from three observables separately, and the confidence contours obtained by combining $\lr{(\delta\pT)^2}$ and $\lr{v_2^2\delta\pT}$ (solid lines). The constraint from $R_{v_2^2}$ is viewed as a lower limit and hence is not used (see main text).}
\vspace*{-0.5cm}
\end{figure*}

Observables in one collision system are strongly influenced by QGP properties during hydrodynamic evolution. By taking ratios between the two systems, such final state effects are largely mitigated: $R_{\mathcal{O}}=\lr{\mathcal{O}}_{\mathrm{U+U}}/\lr{\mathcal{O}}_{\mathrm{Au+Au}}$ (Methods). Figure~\ref{fig:3}a--c show ratios for the three observables. $R_{v_2^2}$ and $R_{(\delta \pT)^2}$ increase by up to 60\% in central collisions, requiring a large $\beta_{2\rm U}$, whereas $R_{v_2^2\delta \pT}$ decreases by up to threefold across centralities, demanding a large $\beta_{2\rm U}$ and a small $\gamma_{\mathrm{U}}$. The ratios in the 0--5\% most central range, having the greatest sensitivity to $^{238}$U shape, are shown as hatch bands in Fig.~\ref{fig:3}d--f.

The data are compared with the state-of-the-art IP-Glasma+MUSIC hydrodynamic model~\cite{Schenke:2020uqq,Schenke:2020mbo}, which combines the fluctuating initial energy density distributions, relativistic viscous hydrodynamics, and hadronic transport. This model, successful in describing flow observables at both RHIC and the LHC~\cite{Schenke:2020mbo}, parameterizes nuclear shapes with a deformed Woods-Saxon profile, 
\begin{align}\label{eq:3}
\rho(r,\theta,\phi) \propto [1+\exp(r-R(\theta,\phi))/a]^{-1}.
\end{align}
The parameters for Au are fixed to an oblate shape with $\beta_{2\mathrm{Au}}=0.14$ and $\gamma_{\mathrm{Au}}=45^{\circ}$~\cite{Bally:2023dxi}, whereas those for U are varied. The model also considers final state effects by adjusting QGP viscosities and initial condition uncertainties, including variations in nuclear radius $R_0$, skin $a$, $\beta_{2\mathrm{Au}}$, $\gamma_{\mathrm{Au}}$, and higher-order shapes (Table~\ref{tab:1} in Methods). These variations are included in the model uncertainties. The mass numbers of Au and U, differing by only 20\%, result in almost completely canceling final state effects, leaving model uncertainties mainly from initial conditions. In Fig.~\ref{fig:3}a--c, calculations for $\beta_{2\mathrm{U}}=0.28$ match central data for $R_{(\delta \pT)^2}$ and $R_{v_2^2\delta \pT}$, but overestimate $R_{v_2^2}$ in 0--30\% centrality. This overestimation stems from the limitations of the model in describing $\varepsilon_2^{\mathrm{rp}}$-induced $v_2$ components, which are strongly affected by variations in the impact parameter and other structural parameters like nuclear radius and skin~\cite{Jia:2022qgl}, as well as possible longitudinal flow decorrelations~\cite{Jia:2024xvl}. Thus, the $R_{v_2^2}$ comparison is expected to set only a lower bound for $\beta_{2\mathrm{U}}$ in this model.

In Fig.~\ref{fig:3}d-f, we contrast the 0--5\% central data against predictions for varying $\beta_{2\mathrm{U}}$ and $\gamma_{\mathrm{U}}$. Calculated $R_{v_2^2}$ and $R_{(\delta \pT)^2}$ change linearly with $\beta_{2\mathrm{U}}^2$, whereas $R_{v_2^2\delta \pT}$ follows a $\beta_{2\mathrm{U}}^3\cos(3\gamma_{\mathrm{U}})$ trend, aligning remarkably with Eq.~\eqref{eq:2}. The intersections between data and model delineate preferred $\beta_{2\mathrm{U}}$ ranges, yielding $\beta_{2\mathrm{U}}(R_{(\delta \pT)^2})=0.294\pm0.021$ and a lower limit value \mbox{$\beta_{2\mathrm{U}}(R_{v_2^2})=0.234\pm0.014$}. For $R_{v_2^2\delta \pT}$, the favored $\beta_{2\mathrm{U}}$ range varies with $\gamma_{\mathrm{U}}$. These preferred ranges are shown in Fig.~\ref{fig:3}g. A combined analysis of constraints from $R_{(\delta \pT)^2}$ and $R_{v_2^2\delta \pT}$ is performed, yielding $\beta_{2\mathrm{U}}=0.297\pm0.015$ and $\gamma_{\mathrm{U}}=8.5^{\circ}\pm4.8^{\circ}$ (mean and one standard deviation, see Methods). 

The data are also compared with Trajectum~\cite{Giacalone:2023cet}, another hydrodynamic model with a different implementation of the initial condition and QGP evolution. Trajectum derives constraints on the initial and final state parameters of the QGP based on a Bayesian analysis of the LHC data that are then extrapolated to the RHIC energies. These constraints were not tuned to the RHIC data but are nevertheless useful in estimating the theoretical uncertainties. The relavent constraints are $\beta_{2\mathrm{U}}=0.275 \pm 0.017$ and $\gamma_{\mathrm{U}}=15.5^{\circ} \pm 7.8^{\circ}$. A combination of constraints from the two models yields $\beta_{2\mathrm{U}}=0.286 \pm 0.025$ and $\gamma_{\mathrm{U}}=8.7^{\circ} \pm 4.5^{\circ}$ (Methods).

The extracted $\beta_{2\mathrm{U}}$ value is in line with low-energy estimates~\cite{Pritychenko:2013gwa}, implying other sources of nucleon, quark and gluon correlations in $^{238}$U are less impactful compared with its large deformation, as supported by recent model studies~\cite{Mantysaari:2023qsq}. Meanwhile, the small $\gamma_{\mathrm{U}}$ value excludes a large triaxiality of Uranium and indicates an average value remarkably consistent with the low energy estimate based on a similar rigid-rotor assumption. This marks the first extraction of nuclear ground state triaxiality without involving transitions to excited states.

{\bf Applications.} The flow-assisted nuclear shape imaging is a promising tool for exploring the structure of atomic nuclei in their ground state. The strength of this method lies in capturing a fast snapshot of nucleon spatial distribution, applicable to any collision species. This contrasts with nuclear spectroscopy, in which complexity varies with the position of nucleus on the Segr\`e chart. This approach is effective for discerning shape differences between species with similar mass numbers, ideally isobar pairs. Many applications are possible, with a few examples given here:
\begin{itemize}
\item Odd-mass nuclides: for odd-mass nuclei, where either $N$ or $Z$ is odd, the nuclear shapes should be similar to adjacent even-even nuclei. As the transition data are more complex~\cite{bohr}, the ground-state shapes are usually inferred from the data measured via laser spectroscopy method~\cite{Yang:2022wbl}. The high-energy approach is suitable for both odd-mass nuclei and even-even nuclei, hence eliminating a possible source of bias in low-energy experiments. 

\item Octupole and hexadecapole deformations: these less common and generally weaker deformations~\cite{RevModPhys.68.349} can be probed through measurement of higher-order flow harmonics (triangular and quadrangular flows). 

\item Dynamic deformations in ``soft'' nuclei: this method could distinguish between average deformation and transient shape fluctuations via measurements of multi-particle correlations~\cite{Dimri:2023wup}. For example, the sixth-particle correlator $\lr{v_2^6}$ has direct sensitivity to the fluctuations of $\gamma$. This information was obtained in rare cases at low energy~\cite{Wu:1996fqm,Ayangeakaa:2019psv}. Our technique, sensitive only to the ground state, also sidesteps the complexities of disentangling shape variations during transitions to excited states.

\item $0\nu\beta\beta$ decay: the decay rate hinges on nuclear matrix elements (NME), significantly affected by the shapes of the initial and final species -- a pair of isobars with the same mass number. Present NME uncertainties, partly stemming from inadequate knowledge of nuclear shapes, pose a main challenge in experimental design~\cite{Engel:2016xgb}. This method, tailored for isobars, allows for precisely determining shape differences between these species. This could reduce NME uncertainties, and hence aid in experiments searching for $0\nu\beta\beta$ decay and enhance our understanding of neutrino properties.
\end{itemize}

It would be remiss not to mention that our approach also holds promise in advancing the study of QGP, particularly its dynamics and transport properties, which have been limited by a poor understanding of QGP initial conditions~\cite{Shuryak:2014zxa,Busza:2018rrf} (Fig.~\ref{fig:1}e). By contrasting flow observables in similarly massed but structurally different species, our technique effectively eliminates final-state effects, thereby isolating initial condition variations seeded by shape differences. This can explain the mechanisms of initial condition formation and consequently help to improve QGP transport property extraction through Bayesian inferences~\cite{Bernhard:2019bmu,JETSCAPE:2020shq,Nijs:2020ors} and lead to breakthroughs in high-energy nuclear physics.

Collective flow assisted nuclear shape imaging is a discovery tool for exploring nuclear structure and high-energy nuclear collision physics. Future research could leverage colliders to conduct experiments with selected isobaric or isobar-like pairs. The combination of high- and low-energy techniques enables interdisciplinary research in the study of atomic nuclei across energy scales.

\subsection{Online content}
Any methods, additional references, Nature Portfolio reporting summaries, source data, extended data, supplementary information, acknowledgements, peer review information; details of author contributions and competing interests; and statements of data and code availability are available at https://doi.org/10.1038/s41586-024-08097-2 .


\section*{Methods}
\subsection{Accessing information in the intrinsic frame}

The nuclear shape in the intrinsic frame is not directly observable in low-energy experiments. However, in high-energy collisions, the collective flow phenomenon is sensitive to the shape and size of the nucleon distribution in the overlap region of the transverse plane. This distribution, denoted as $\rho\left(\mathbf{r}\right)$ with ${\bf r} = x+iy$, provides a direct link to the shape characteristics of the two colliding nuclei in their intrinsic frames, as discussed below.

The elliptic shape of the heavy-ion initial state is characterized by its amplitude $\varepsilon_2$ and direction $\Phi_2$, defined by nucleon positions as,
\begin{align}\label{eq:a1}
\mathcal{E}_2 \equiv \varepsilon_2 e^{2i\Phi_2} = \frac{\int_{\mathbf{r}}\mathbf{r}^2 \rho\left(\mathbf{r}\right)}{\int_{\mathbf{r}}|\mathbf{r}|^2\rho(\mathbf{r})}, \int_{\mathbf{r}}=\int \mathrm{d} x \mathrm{d} y\;.
\end{align}
When the coordinate system is rotated such that $x$ and $y$ coincide with the minor and major axes, the elliptic eccentricity coincides with the usual definition $\varepsilon_2=\frac{\lr{y^2}-\lr{x^2}}{\lr{y^2}+\lr{x^2}}$. The parameter $\varepsilon_2$ drives the elliptic flow $v_2$: $v_2\propto \varepsilon_2$.

Let's now consider collisions at zero impact parameter, where, without loss of generality, the average elliptic geometry vanishes, i.e., $\lr{\mathcal{E}_2}=0$. The second moment of eccentricity over many events is given by~\cite{Giacalone:2019kgg,Giacalone:2023hwk},
\small{\begin{align}\label{eq:a2}
\lr{\varepsilon_2^2} = \lr{\mathcal{E}_2\mathcal{E}_2^*}\approx \frac{\int_{\mathbf{r}_1, \mathbf{r}_2}\left(\mathbf{r}_1\right)^2\left(\mathbf{r}_2^*\right)^2 \rho\left(\mathbf{r}_1, \mathbf{r}_2\right)}{\left(\int_{\mathbf{r}}|\mathbf{r}|^2\lr{\rho(\mathbf{r})}\right)^2}\;,
\end{align}}\normalsize
where $\lr{\rho(\mathbf{r})}$ represents the event-averaged profile, and 
\small{\begin{align}\nonumber
 \rho\left(\mathbf{r}_1, \mathbf{r}_2\right) =\lr{\delta\rho(\mathbf{r}_1)\delta\rho(\mathbf{r}_2)} = \lr{\rho(\mathbf{r}_1)\rho(\mathbf{r}_2)}-\lr{\rho(\mathbf{r}_1)}\lr{\rho(\mathbf{r}_2)}
\end{align}}\normalsize
is the usual two-body distribution. Similarly, the third central moments are related to the three-body distribution, $\rho\left(\mathbf{r}_1, \mathbf{r}_2, \mathbf{r}_3\right)=\lr{\delta\rho(\mathbf{r}_1)\delta\rho(\mathbf{r}_2)\delta\rho(\mathbf{r}_3)}$. For example,
\begin{align}\label{eq:a4}
 \lr{\varepsilon_2^2 \delta d_{\perp}/d_{\perp}} &\approx -\frac{\int_{\mathbf{r}_1, \mathbf{r}_2,\mathbf{r}_3}\left(\mathbf{r}_1\right)^2\left(\mathbf{r}_2^*\right)^2 |\mathbf{r}_3^2|\rho\left(\mathbf{r}_1, \mathbf{r}_2, \mathbf{r}_3\right)}{\left(\int_{\mathbf{r}}|\mathbf{r}|^2\lr{\rho(\mathbf{r})}\right)^3}\;,
\end{align}
where we define $\delta d_{\perp}/d_{\perp}\equiv (d_{\perp}-\lr{d_{\perp}})/\lr{d_{\perp}}$, and the relation $\frac{\delta d_{\perp}}{d_{\perp}} \approx -\frac{\delta\lr{|\mathbf{r}^2|}}{\lr{|\mathbf{r}^2|}} =  -\frac{\int_{\mathbf{r}} |\mathbf{r}^2| \delta\rho\left(\mathbf{r}\right)}{\int_{\mathbf{r}}|\mathbf{r}|^2\langle\rho(\mathbf{r})\rangle}$ is used.

The quantities $\mathcal{E}_2$ and $\delta d_{\perp}/d_{\perp}$ depend not only on the nuclear shape but also the random orientations of the projectile and target nuclei, denoted by Euler angles $\Omega_p$ and $\Omega_t$. For small quadrupole deformation, it suffices to consider the leading-order forms~\cite{Jia:2021qyu}:
\begin{align}\nonumber
\frac{\delta d_{\perp}}{d_{\perp}} &\approx  \delta_d + p_0(\Omega_p,\gamma_p)\beta_{2p}+p_0(\Omega_t,\gamma_t)\beta_{2t}\;,\\\label{eq:a5}
\mathcal{E}_2 &\approx \mathcal{E}_0 + {\bm p}_{2}(\Omega_p,\gamma_p)\beta_{2p} +{\bm p}_{2}(\Omega_t,\gamma_t)\beta_{2t}.
\end{align}
Here, the scalar $\delta_d$ and vector $\mathcal{E}_0$ represent values for spherical nuclei. The values of scalar $p_0$ and vector ${\bm p}_{2}$ are directly connected to the $xy$-projected one-body distribution $\rho(\mathbf{r})$. Therefore, they depend on the orientation of the two nuclei.  The fluctuations of $\delta_d$ ($\mathcal{E}_0$) are uncorrelated with $p_0$ (${\bm p}_{2}$). After averaging over collisions with different Euler angles and setting $\beta_{2p}=\beta_{2t}$ and $\gamma_p=\gamma_t$, we obtain: 
\begin{align}\nonumber
\lr{\varepsilon_2^2} &= \lr{\varepsilon_0^2} +  2\lr{{\bm p}_{2}(\gamma){\bm p}_{2}^*(\gamma)}\beta_2^2\\\nonumber
\lr{(\delta d_{\perp}/d_{\perp})^2} &= \lr{\delta_d^2} +  2\lr{p_{0}(\gamma)^2}\beta_2^2\\\label{eq:a6}
\lr{\varepsilon_2^2 \delta d_{\perp}/d_{\perp}} &=  \lr{\varepsilon_0^2\delta_d} +2\lr{p_0(\gamma){\bm p}_{2}(\gamma){\bm p}_{2}(\gamma)^*}\beta_2^3\;.
\end{align}
It is found that $\lr{{\bm p}_{2}(\gamma){\bm p}_{2}^*(\gamma)}$ and $\lr{p_{0}(\gamma)^2}$ are independent of $\gamma$, while $\lr{p_0(\gamma){\bm p}_{2}(\gamma){\bm p}_{2}(\gamma)^*}\propto -\cos (3\gamma)$, resulting in expressions in Eq.~\eqref{eq:2}. 

The event-averaged moments in Eq.~\eqref{eq:a6} are rotationally invariant and capture the intrinsic many-body distributions of $\rho\left(\mathbf{r}\right)$. Note that the coefficients $a_n$ in Eq.~\eqref{eq:2} are strong functions of centrality that decrease towards central collisions, while coefficients $b_n$ vary weakly with centrality. Therefore, the impact of deformation is always largest in the most central collisions. In general, it can be shown that the $n$-particle correlations reflect the rotational invariant ${n}^{\mathrm{th}}$ central moments of $\rho\left(\mathbf{r}\right)$, which in turn are connected to the ${n}^{\mathrm{th}}$ moments of the nuclear shape in the intrinsic frame. 

\subsection{Previous experimental attempts on nuclear shapes at high energy}

The idea that $v_2$ can be enhanced by $\beta_2$ was recognized early~\cite{Li:1999bea,Shuryak:1999by,Filip:2009zz,Shou:2014eya,Goldschmidt:2015kpa}. Studies at RHIC~\cite{Adamczyk:2015obl} and the LHC~\cite{Acharya:2018ihu,Sirunyan:2019wqp,Aad:2019xmh} in $^{238}$U+$^{238}$U and $^{129}$Xe+$^{129}$Xe collisions indicated $\beta_2$'s influence on $v_2$. Several later theoretical investigations assessed the extent to which $\beta_2$ can be constrained by $v_2$ alone~\cite{Giacalone:2017dud,Giacalone:2020awm,Giacalone:2021udy,Ryssens:2023fkv}. A challenge with $v_2$ is that its $a_1$ term in Eq.~\eqref{eq:2} is impacted by $\varepsilon_2^{\mathrm{rp}}$, which often exceeds the $b_1\beta_2^2$ term even in central collisions. A recent measurement of $\lr{v_2^2\delta \pT}$ aimed to assess the triaxiality of $^{129}$Xe~\cite{ATLAS:2022dov}, but the extraction of $\gamma_{\mathrm{Xe}}$ was hindered by needing prior knowledge of $\beta_{2\mathrm{Xe}}$ and potentially substantial fluctuations in $\gamma_{\mathrm{Xe}}$~\cite{Robledo:2008zz,Bally:2021qys,Bally:2022rhf,Dimri:2023wup,Fortier:2023xxy}. The combination of several observables in this article allows for more quantitative extraction of nuclear shape parameters.

\subsection{Event selection} 
In high-energy experiments, the polar angle $\theta$ is usually mapped to the so-called pseudorapidity variable $\eta= -\ln(\tan(\theta/2))$. The STAR TPC polar angle range $|\theta-90^{\circ}|<50^{\circ}$ corresponds to $|\eta|<1$.

The collision events are selected by requiring a coincidence of signals from two vertex position detectors situated on each side of the STAR barrel, covering a pseudorapidity range of $4.4 < |\eta| < 4.9$. To increase the statistics for ultra-central collision (UCC) events, a special sample of Au+Au data in 2010 and U+U data is chosen based on the criteria of high multiplicity in the STAR TPC and minimal activity in the zero-degree calorimeters that cover the beam rapidity~\cite{Bieser:2002ah}. 

In the offline analysis, events are selected to have collision vertices $z_{\mathrm{vtx}}$ within 30 cm of the TPC center along the beamline and within 2 cm of the beam spot in the transverse plane. Additionally, a selection criterion based on the correlation between the number of TPC tracks and the number of tracks matched to the time-of-flight detector covering $|\eta|<0.9$ is applied to suppress pileup events (events containing more than one collision in the TPC)~\cite{Llope:2012zz} and background events. 

After applying these selection criteria, the Au+Au dataset has approximately 528 million minimum-bias events (including 370 million in 2011) and 120 million UCC events. The U+U dataset comprises around 300 million minimum-bias events and 5 million UCC events.

\subsection{Track selection}
For this analysis, tracks are selected with $|\eta|<1$ and the transverse momentum range $0.2 < \pT < 3.0 $ GeV/$c$. To ensure good quality, the selected tracks must have at least 16 fit points out of a maximum of 45, and the ratio of the number of fit points to the number of possible points must be greater than 0.52. Additionally, to reduce contributions from secondary decays, the distance of the track's closest approach (DCA) to the primary collision vertex must be less than 3 cm. 

The tracking efficiency in the TPC was evaluated using the standard STAR Monte Carlo embedding technique~\cite{STAR:2015tnn}. The efficiencies are nearly independent of $\pT$ for $\pT>0.5$ GeV/$c$, with plateau values ranging from 0.72 (0.69) in the most central Au+Au (U+U) collisions to 0.92 in the most peripheral collisions. The efficiency exhibits some $\pT$-dependent variation, on the order of 10\% of the plateau values, within the range of $0.2<\pT<0.5$ GeV/$c$.

\subsection{Centrality}
The centrality of each collision is determined using $\nchrec$, which represents the number of raw reconstructed tracks in $|\eta|<0.5$, satisfying $\pT> 0.15$ GeV/$c$, and having more than 10 fit points. After applying a correction to account for the dependence on the collision vertex position and the luminosity, the distribution of $\nchrec$ is compared with a Monte Carlo Glauber calculation~\cite{STAR:2015tnn}. This comparison allows for determining centrality intervals, expressed as a percentage of the total nucleus-nucleus inelastic cross-section. 

\subsection{Calculation of observables} 
The $\lr{v_2^2}$, $\lr{(\delta \pT)^2}$, and $\lr{v_2^2\delta \pT}$ are calculated using charged tracks as follows:
\scriptsize{\begin{align}\nonumber
[\pT] &= \frac{\sum_{i}w_ip_{\mathrm{T},i}}{\sum_{i}w_i},\llrr{\pT} \equiv \lr{[\pT]}_{\mathrm{evt}} \\\nonumber
\lr{(\delta \pT)^2}&=\lr{\frac{\sum_{i\neq j}w_iw_j(p_{\mathrm{T},i}-\llrr{\pT})(p_{\mathrm{T},j}-\llrr{\pT})}{\sum_{i\neq j}w_iw_j}}_{\mathrm{evt}}\\\nonumber
\lr{v_2^2} & =  \lr{ \frac{\sum_{i\neq j}w_iw_j \cos (2(\phi_i-\phi_j))}{\sum_{i\neq j}w_iw_j}}_{\mathrm{evt}}\\\label{eq:a7}
\lr{v_2^2\delta \pT} &= \lr{\frac{\sum_{i\neq j\neq k}w_iw_jw_k\cos (2(\phi_i-\phi_j))(p_{\mathrm{T},k}-\llrr{\pT})}{\sum_{i\neq j\neq k}w_iw_jw_k}}_{\mathrm{evt}}\;.
\end{align}}\normalsize
The averages are performed first on all multiplets within a single event and then over all events in a fixed $\nchrec$ bin. The track-wise weights $w_{i,j,k}$ account for tracking efficiency and its $\eta$ and $\phi$ dependent variations. The values of $\lr{v_2^2}$ and $\lr{(\delta \pT)^2}$ are obtained using the standard method, in which particle $i$ and $j$ are selected from $|\eta|<1$, as well as the two-subevent method, in which particle $i$ and $j$ are selected from pseudorapidity ranges of $-1 < \eta_i < -0.1$ and $0.1 < \eta_j < 1$, respectively. We also calculate the efficiency-corrected charged particle multiplicity in $|\eta|<0.5$, defined as $\nch = \sum_{i}w_i$. This observable is used for evaluating the systematics. 

The covariance $\lr{v_2^2\delta \pT}$ is calculated by averaging over all triplets labeled by particle indices $i$, $j$, and $k$. The standard cumulant framework is used to obtain the results instead of directly calculating all triplets~\cite{Bilandzic:2010jr}. We also calculated $\lr{v_2^2\delta \pT}$ using the two-subevent method~\cite{ATLAS:2022dov}, where particles $i$ and $j$ are taken from ranges of $-1 < \eta_i < -0.1$ and $0.1 < \eta_j < 1$, while particle $k$ is taken from either subevents. Including a pseudorapidity gap between the particle pairs or triplets suppresses the short-range ``non-flow'' correlations arising from resonance decays and jets~\cite{Jia:2017hbm}. 

\begin{figure}[!h]
\begin{center}
\includegraphics[width=1\linewidth]{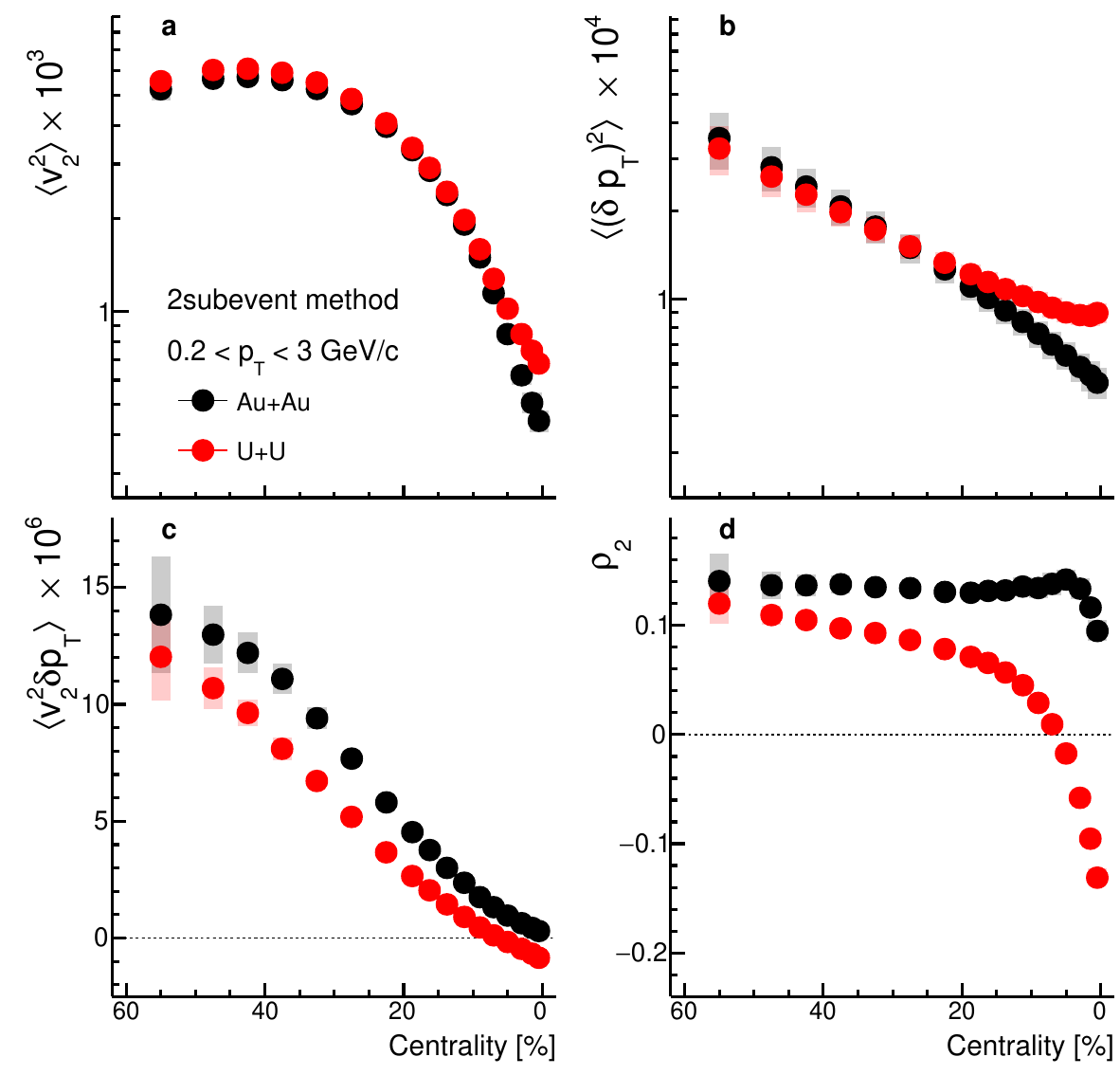}
\end{center}
\caption{\label{fig:a1} \textbf{Components involved in $v_2$-$[\pT]$ correlations}. The centrality dependences of $\lr{v_2^2}$ (\textbf{a}), $\lr{(\delta \pT)^2}$ (\textbf{b}), $\lr{v_2^2\delta \pT}$ (\textbf{c}) and $\rho_2~=~\frac{\lr{v_2^2 \delta \pT}}{\lr{v_2^2}\sqrt{\lr{(\delta \pT)^2}}}$ (\textbf{d}) in U+U (red) and Au+Au (black) collisions. They are calculated using the two-subevent method. The error bars and shaded bands are statistical and systematic uncertainties, respectively.}\vspace*{-0.5cm}
\end{figure}

The calculation of $\rho_2~=\frac{\lr{v_2^2 \delta \pT}}{\lr{v_2^2}\sqrt{\lr{(\delta \pT)^2}}}$ relies on the input values of $\lr{v_2^2}$, $\lr{(\delta \pT)^2}$ and $\lr{v_2^2\delta \pT}$. These components and $\rho_2$ are shown in Fig.~\ref{fig:a1} as a function of centrality. In the central region, enhancements of $\lr{v_2^2}$ and $\lr{(\delta \pT)^2}$ are observed in U+U relative to Au+Au collisions, which is consistent with the influence of large $\beta_{2\rm U}$. In contrast, the values of $\lr{v_2^2\delta \pT}$ are significantly suppressed in U+U compared with Au+Au collisions across the entire centrality range shown. This suppression is consistent with the negative contribution expected for strong prolate deformation of U as described in Eq.~\eqref{eq:2}. 

In this analysis, the default results are obtained using the two-subevent method. The differences between the standard and two-subevent methods are used to evaluate the impact of non-flow correlations discussed below.

\subsection{Influence of non-flow correlations} An important background in our measurement is non-flow: correlations among a few particles originated from a common source, such as resonance decays and jets, which are uncorrelated with the initial geometry. Two approaches are used to estimate the non-flow contributions. Non-flow correlations are short-range in $\eta$ and can be suppressed via the subevent method by requiring a rapidity gap between the pairs or triplets of particles in Eq.\eqref{eq:a7}. Hence, in the first approach, the differences between the standard and subevent methods provide an estimate of the non-flow contribution. However, one must keep in mind that part of the rapidity gap dependence of the signal in central collisions may arise from longitudinal fluctuations in $[\pT]$ and $v_2$ due to variations in the initial geometry in $\eta$~\cite{ATLAS:2022dov}. 

The second approach assumes that the clusters causing non-flow correlations are mutually independent. In this independent-source scenario, non-flow in $n$-particle cumulants is expected to be diluted by the charged particle multiplicity as $1/\nch^{n-1}$~\cite{Borghini:2000sa}. Therefore, non-flow (nf) contributions in a given centrality can be estimated by, 
\begin{align}\nonumber
&\lr{v_2^2}_{\mathrm{nf}} \approx \frac{\left[\lr{v_2^2}\nch\right]_{\mathrm{peri}}}{\nch} \;,\\\nonumber
&\lr{(\delta \pT)^2}_{\mathrm{nf}} \approx \frac{\left[\lr{(\delta \pT)^2}\nch\right]_{\mathrm{peri}}}{\nch}\;, \\\label{eq:a7b}
&\lr{v_2^2\delta \pT }_{\mathrm{nf}} \approx \frac{\left[\lr{v_2^2\delta \pT }\nch^2\right]_{\mathrm{peri}}}{\nch^2} 
\end{align}
where the subscript ``peri'' is a label for the peripheral bin. This procedure makes two assumptions that are not fully valid: 1) the signal in the peripheral bin is all non-flow, and 2) non-flow in other centralities is unmodified by final state medium effects.  For example, the medium effects strongly suppress the jet yield and modify the azimuthal structure of non-flow correlations. Hence, this approach only provides a qualitative estimate of the non-flow. Moreover, this approach is not applicable for $\lr{(\delta\pT)^2}$, as medium effects are expected to reduce the momentum differences of non-flow particles, since they are out of local equilibrium. 

\begin{figure*}[!htb]
\begin{center}
\includegraphics[width=0.8\linewidth]{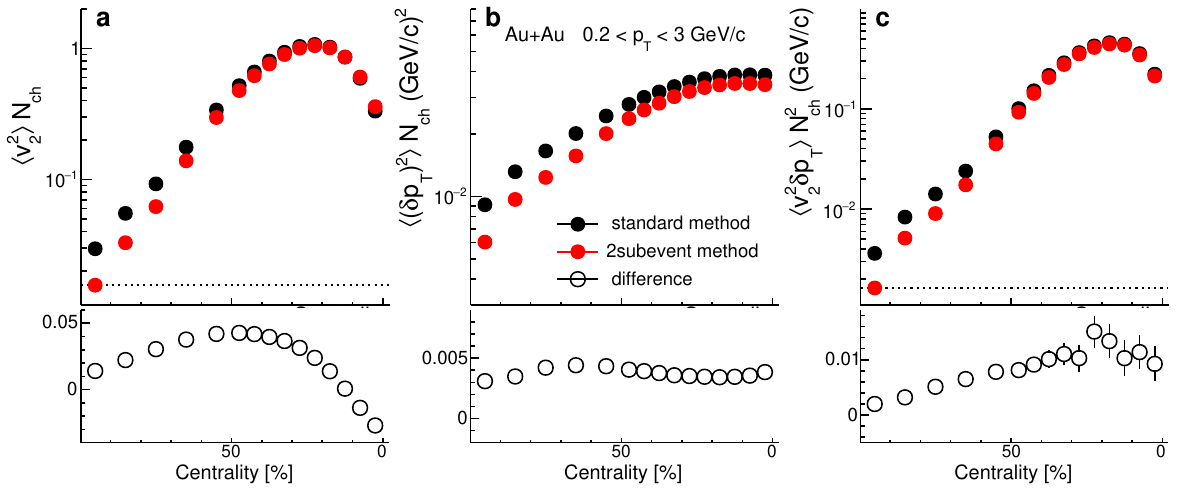}
\end{center}
\caption{\label{fig:a2} \textbf{Impact of non-flow correlations}. The centrality dependence of $\nch$ scaled quantities in Au+Au collisions, $\lr{v_2^2}\nch$ (\textbf{a}), $\lr{(\delta \pT)^2}\nch$ (\textbf{b}), and $\lr{v_2^2 \delta \pT}\nch^2$ (\textbf{c}) from the standard method (black solid markers), two-subevent method (red solid markers) and their differences (open markers). The error bars are statistical uncertainties. The decrease of $\lr{v_2^2}$-difference towards central collisions is attributed to two track reconstruction effects in the standard method, which does not affect the other two observables.  The dashed lines indicate the upper limit of the non-flow contribution, assuming $\nch$-scaling of Eq.~\eqref{eq:a7b}.}
\end{figure*}

Figure~\ref{fig:a2} shows the $\nch$-scaled values of $\lr{v_2^2}$, $\lr{(\delta \pT)^2}$, and $\lr{v_2^2\delta \pT}$ as a function of centrality in Au+Au collisions. The requirement of subevent reduces the signal in the most peripheral bin by 50\%, 40\%, and 80\%, respectively, which can be treated as the amount of non-flow rejected by the subevent requirement. Therefore, we use the differences between the standard and subevent methods to estimate the non-flow in the subevent method. These differences vary with centrality because of the combined effects of medium modification of non-flow and longitudinal flow decorrelations~\cite{ATLAS:2020sgl}. These differences are propagated to the ratios of these observables between U+U and Au+Au. They are found to be 1.1\%, 3.5\%, and 11\% for  $R_{v_2^2}$, $R_{(\delta \pT)^2}$, and $R_{v_2^2\delta \pT}$, respectively. 

\begin{figure*}[!htb]
\begin{center}
\includegraphics[width=0.8\linewidth]{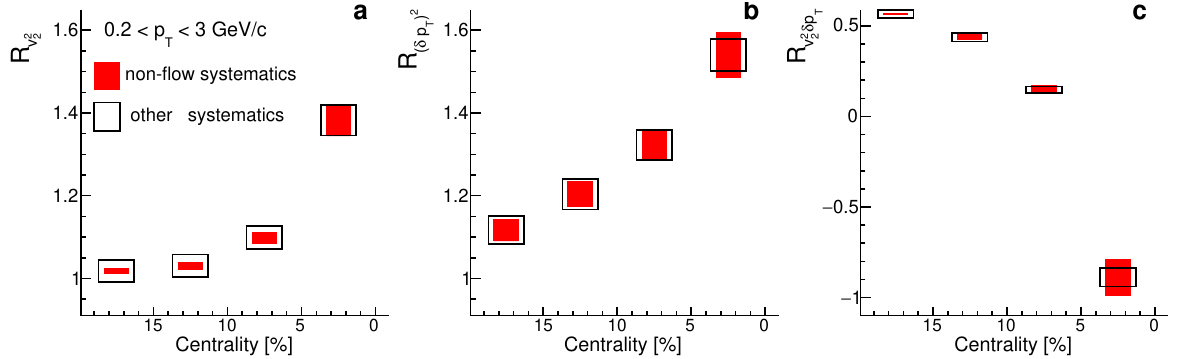}
\end{center}
\caption{\label{fig:a3} \textbf{Summary of the systematic uncertainties}. The centrality dependence of ratios $R_{v_2^2}$ (a), $R_{(\delta \pT)^2}$ (b), and $R_{v_2^2 \delta \pT}$ (c) from the two-subevent method with the systematic uncertainties arising from the non-flow estimate (red bars) and other sources (black boxes).}
\end{figure*}

Figure~\ref{fig:a2} also provides an estimate of non-flow based on the $\nch$-scaling method. We assume that the entire signals in the 80--100\% centrality in two-subevent are non-flow, and then use  Eq.~\eqref{eq:a7b} to estimate the fraction of non-flow as a function of centrality. As mentioned earlier, we employ this approach for $\lr{v_2^2}$ and $\lr{v_2^2\delta \pT}$, where the medium effects may redistribute non-flow correlations in azimuthal angle, instead of suppressing them. This approach is unsuitable for $\lr{(\delta \pT)^2}$, for which the medium effects should always suppress the non-flow contribution. In the 0-5\% most central collisions, the estimated non-flow is about 6\% for $\lr{v_2^2}$ and only about 1.4\% for $\lr{v_2^2\delta \pT}$. These differences, when propagated to the ratios, are reduced for $R_{v_2^2}$, which is positive, and increased for $R_{v_2^2\delta \pT}$, which is negative. They amount to about 2.8\% for $R_{v_2^2}$ and 2.5\% for $R_{v_2^2\delta \pT}$, respectively.

The non-flow systematic uncertainties are taken as the larger of the two approaches for $R_{v_2^2}$ and $R_{v_2^2\delta \pT}$, while for $R_{(\delta \pT)^2}$, the difference between standard and subevent methods is used. The total non-flow uncertainties in the 0--5\% centrality are 2.8\%, 3.5\% and 11\% for $R_{v_2^2}$, $R_{(\delta \pT)^2}$ and $R_{v_2^2\delta \pT}$, respectively.  

Figure~\ref{fig:a3} contrasts the non-flow systematic uncertainties with other sources of uncertainties (next section) in this analysis. In the 0--5\% centrality, the non-flow uncertainties are comparable or somewhat larger than other sources, while they are subdominant in other centrality ranges. 

In the literature, non-flow contributions are sometimes estimated using the HIJING model~\cite{Gyulassy:1994ew}, which has only non-flow correlations. The latter were found to follow very closely Eq.~\eqref{eq:a7b}~\cite{Bhatta:2021qfk,Wang:2024zml}. In our second approach, instead of relying on the HIJING model, we assume this $\nch$-scaling behavior but use real peripheral data as the baseline for non-flow contributions. Our findings indicate that the HIJING model fails to quantitatively capture the features of non-flow. Specifically, HIJING predicts a much weaker $\Delta\eta$ dependence for $\lr{v_2^2}$, with only a 13\% difference between the standard and two-subevent method, whereas the data indicate a 50\% decrease~\cite{STAR:2023wmd} (Fig. 25 in Ref.~\cite{STAR:2023wmd} for $p+p$ collisions). Furthermore, we found that the values of $\lr{v_2^2\delta \pT}$ predicted by HIJING are three times larger than the data in peripheral Au+Au collisions. Therefore, the non-flow estimation based on the HIJING model in Ref.~\cite{Wang:2024zml} appears to be exaggerated. A more recent estimate~\cite{Jia:2024xvl}, based on a transport model incorporating full medium dynamics and Eq.~\eqref{eq:a7b}, yields a non-flow fraction consistent with STAR data. This study also suggests a potential increase of the $R_{v_2^2}$ with $\Delta\eta$ associated with flow decorrelation effects.

Understanding non-flow correlations as a physical process has always been a work in progress. As our knowledge deepens, the non-flow uncertainties are expected to reduce. Rather than merely contributing to experimental uncertainties or even being corrected for in the data, non-flow physics should ultimately be incorporated into hydrodynamic models. Currently, these models include non-flow effects from resonance decays but lack contributions from jet fragmentation.

\subsection{Systematic uncertainties} Systematic uncertainties include an estimate of the non-flow contributions discussed above and other sources accounting for detector effects and analysis procedure.  These other sources are estimated by varying the track quality selections, the $z_{\mathrm{vtx}}$ cuts, examining the influence of pileup, comparing results from periods with different detector conditions, and closure test. The influence of track selection criteria is studied by varying the number of fit hits on the track from a minimum of 16 to 19 and by varying DCA cut from $<3$ cm to $<2.5$ cm, resulting in variations of 1--5\% for $\lr{(\delta \pT)^2}$. The impacts on $\lr{v_2^2}$ and $\lr{v_2^2\delta\pT}$ are up to $2.5\%$ and $4\%$, respectively. 

The influence of track reconstruction on the collision vertex is examined by comparing the results for different $|z_{\mathrm{vtx}}|$ cuts, with variations found to be $0.5$--3\% for all observables. Comparisons between data-taking periods, particularly normal and reverse magnetic field runs in Au+Au collisions, show consistency within their statistical uncertainties. The influence of pileup and background events is studied by varying the cut on the correlation between $\nchrec$ and the number of hits in the TOF. The influence is found to be 1--3\% for $\lr{v_2^2}$ and $\lr{(\delta \pT)^2}$, and reaches 2--10\% for $\lr{v_2^2\delta \pT}$. Comparisons are also made between the 2010 and 2011 Au+Au datasets, which have different active acceptances in the TPC. The results are largely consistent with the quoted uncertainties, although some differences are observed, particularly in the central region, where variations reach 5--10\% for $\lr{v_2^2\delta \pT}$. 

A closure test was conducted, wherein the reconstruction efficiency and its variations in $\eta$ and $\phi$ from the data were utilized to retain a fraction of the particles generated from a multi-phase transport model~\cite{Lin:2004en}. Subsequently, a track-by-track weight, as described in Eq.~\eqref{eq:a7}, was applied to the accepted particles. All observables are calculated using the accepted particles and compared with those obtained using the original particles. This procedure allowed us to recover $\lr{v_2^2}$ and $\lr{(\delta \pT)^2}$ within their statistical uncertainties. However, a 2--3\% nonclosure was observed in $\lr{v_2^2\delta \pT}$. Nevertheless, it's important to note that such non-closures largely cancel when considering the ratios between U+U and Au+Au collisions.

Several additional cross-checks were carried out. The track reconstruction efficiency has an approximate 5\% uncertainty due to its reliance on particle type and occupancy dependence. We repeated the analysis by varying this efficiency, and the variations in the results were either less than 1\% or consistent within their statistical uncertainties. The reconstructed $\pT$ can differ from the true value due to finite momentum resolution. This effect was investigated by smearing the reconstructed $\pT$ according to the known resolution, calculating the observable, and comparing the results with the original ones. A discrepancy of approximately 0.5\% was observed for $\lr{(\delta \pT)^2}$, while other observables remained consistent within their statistical uncertainties. These effects cancel in the ratios between U+U and Au+Au collisions.

The default results are obtained from the two-subevent method. The total systematic uncertainties, including these sources and non-flow, are calculated as a function of centrality. The uncertainties of the ratios between U+U and Au+Au are evaluated for each source and combined in quadrature to form the total systematic uncertainties.  This process results in a partial cancellation of the uncertainties between the two systems.  The uncertainties from different sources discussed above on the ratios are shown by the black boxes in Fig.~\ref{fig:a3}. The total systematic uncertainties including non-flow in the 0--5\% centrality range amount to 3.9\%, 4.4\%, and 12.5\% for $R_{v_2^2}$, $R_{(\delta \pT)^2}$, and $R_{v_2^2\delta \pT}$, respectively. 

\subsection{Hydrodynamic model setup and simulation} Table~\ref{tab:1} details the Woods-Saxon parameters for Au and U used in the IP-Glasma+MUSIC model calculations. The nucleon-nucleon inelastic cross-sections are the standard values 42 mb and 40.6 mb for Au+Au collisions at 200 GeV and U+U collisions at 193 GeV, respectively. For U, the nuclear shape in Eq.~\eqref{eq:1} is extended to include a possible small axial hexadecapole deformation $\beta_4$:
\small{\begin{align}\label{eq:a9}
R(\theta,\phi) &= R_0(1+\beta_2 [\cos \gamma Y_{2,0}+ \sin\gamma Y_{2,2}]+\beta_4  Y_{4,0})\;.
\end{align}}\normalsize

Most low-energy nuclear structure models favor a modest oblate deformation for $^{197}$Au~\cite{Bally:2023dxi}. We assume $\beta_{2\mathrm{Au}}=0.14$ and and $\gamma_{\mathrm{Au}}=45^{\circ}$ as the default choice for $^{197}$Au, which are varied within the range in the range of $\beta_{2\mathrm{Au}}\approx 0.12-0.14$ and $\gamma_{\mathrm{Au}}\approx 37-53^{\circ}$ according to Refs.~\cite{Ryssens:2023fkv,Bally:2023dxi}. These calculations reasonably reproduce many observables related to the ground-state nuclear deformation. For $^{238}$U, we scan several $\beta_{2\mathrm{U}}$ values ranging from 0 to 0.34. We also vary $\beta_{4\mathrm{U}}$ from 0 to 0.09 and $\gamma_{\mathrm U}$ in the range of 0$^{\circ}$--20$^{\circ}$ to examine the sensitivity of the U+U results to hexadecapole deformation and triaxiality.  For each setting, about 100,000--400,000 events are generated using the officially available IP-Glasma+MUSIC~\cite{Schenke:2020uqq,Schenke:2020mbo}. Each event is oversampled at least 100 times to minimize statistical fluctuations in the hadronic transport. These calculations were performed using services provided by the Open Science Grid (OSG) Consortium~\cite{osg07,osg09}.

\begin{figure*}[!htb]
\begin{center}
\includegraphics[width=0.8\linewidth]{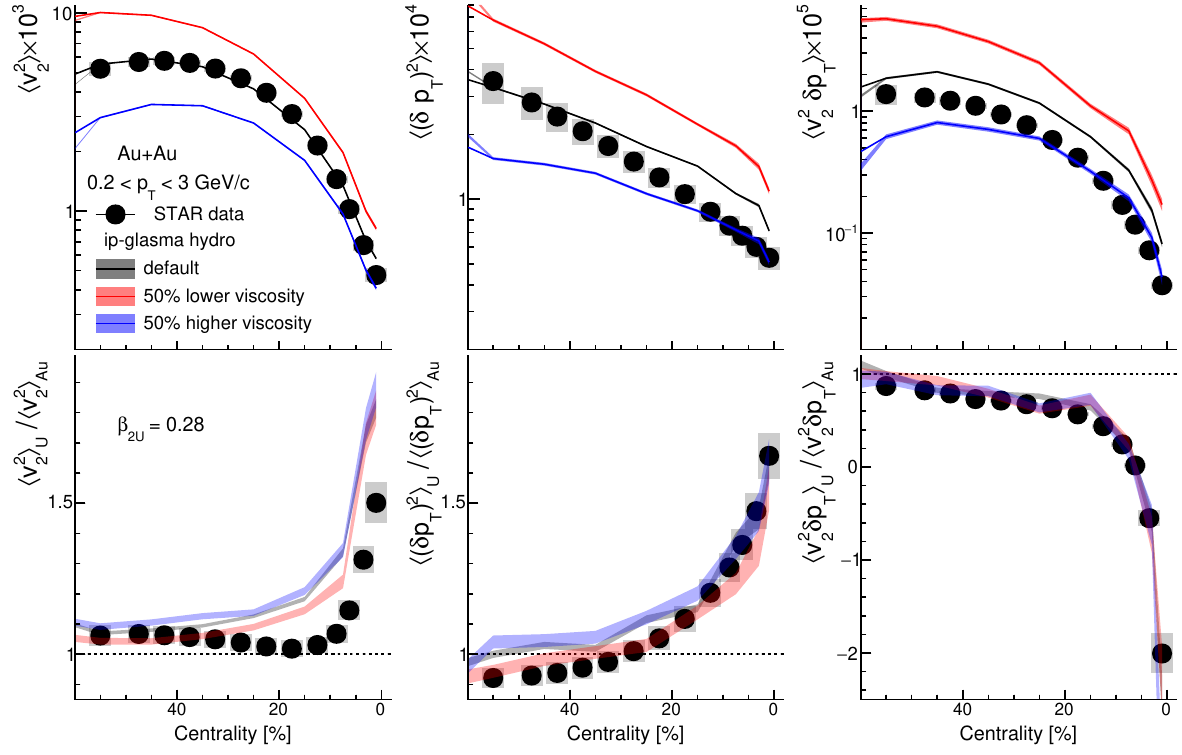}
\end{center}
\caption{\label{fig:a4} \textbf{Sensitivity to shear and bulk viscosities of the QGP}. IP-Glasma+MUSIC model prediction of the $\lr{v_2^2}$ (left), $\lr{(\delta \pT)^2}$ (middle) and $\lr{v_2^2\delta \pT}$ (right) compared with the data in Au+Au collisions and the corresponding ratios (bottom) between U+U and Au+Au collisions, for different amount of shear and bulk viscosities.}
\end{figure*}

The role of final state effects is studied by varying the shear and bulk viscosities simultaneously up and down by 50\%. The impacts on $\lr{v_2^2}$, $\lr{(\delta\pT)^2}$, and $\lr{v_2^2\delta\pT}$ are shown for Au+Au collisions in top of panels of Fig.~\ref{fig:a4}. The values of these flow observables are changed by more than a factor of two as a function of centrality. Yet, the ratios between U+U and Au+Au collisions, shown in the bottom panels, are relatively stable. A small reduction of $R_{v_2^2}$ and $R_{(\delta \pT)^2}$ are observed in non-central collisions, when values of viscosities are halved.  However, this change is an overestimate since the calculated flow observables greatly overestimate the data. So, in the end, half of the variations of the ratios are included in the model uncertainty. 

The main theoretical uncertainties arise from variations in nuclear structure parameters.  Parameters common between two collision systems, such as the minimum inter-nucleon distance in nuclei $d_{\mathrm {min}}$, are not expected to contribute to the uncertainty significantly. However, other parameters, including nuclear radius $R_{0}$, skin $a$, and higher-order hexadecapole deformation $\beta_4$, could be different between Au and U and hence contribute more to the theoretical uncertainty. 

\begin{figure*}[!htb!]
\begin{center}
\includegraphics[width=0.8\linewidth]{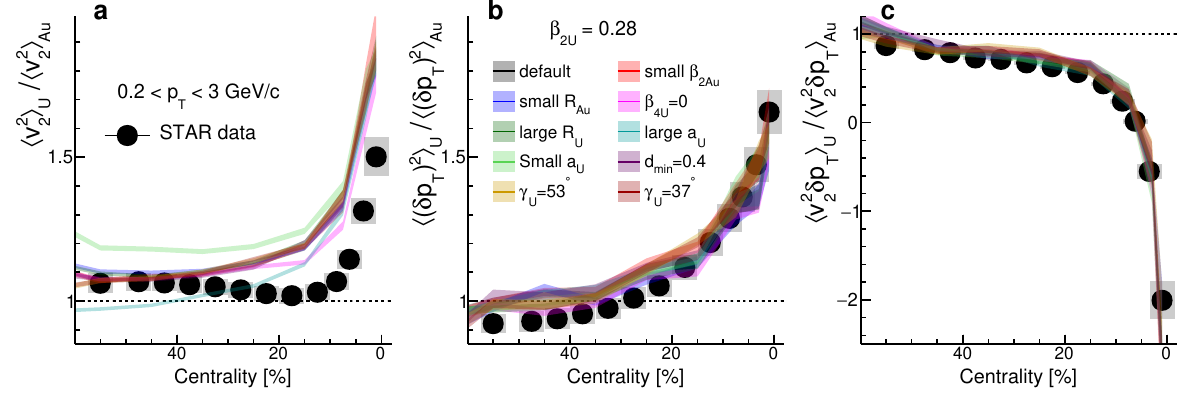}
\end{center}
\caption{\label{fig:a5} \textbf{Sensitivity to nuclear structure parameters}. IP-Glasma+MUSIC model prediction of the ratios of $\lr{v_2^2}$ (\textbf{a}), $\lr{(\delta \pT)^2}$ (\textbf{b}), and $\lr{v_2^2\delta \pT}$ (\textbf{c}) between U+U and Au+Au collisions. The calculations are done for different Glauber model parameters, and they are compared with the data.}
\end{figure*}

Table~\ref{tab:1} provides a list of variations of nuclear structure parameters. The impact of these variations on ratios of flow observables is displayed in Fig.~\ref{fig:a5}. The ratios of flow observables are insensitive to these variations in the most central collisions. $R_{v_2^2}$ is particularly sensitive to skin parameter $a$. This is understandable, as $v_2$ has a large contribution from the reaction plane flow, which varies strongly with the value of $a$~\cite{Jia:2022qgl}.
\begin{figure*}[!htb!]
\begin{center}
\includegraphics[width=0.8\linewidth]{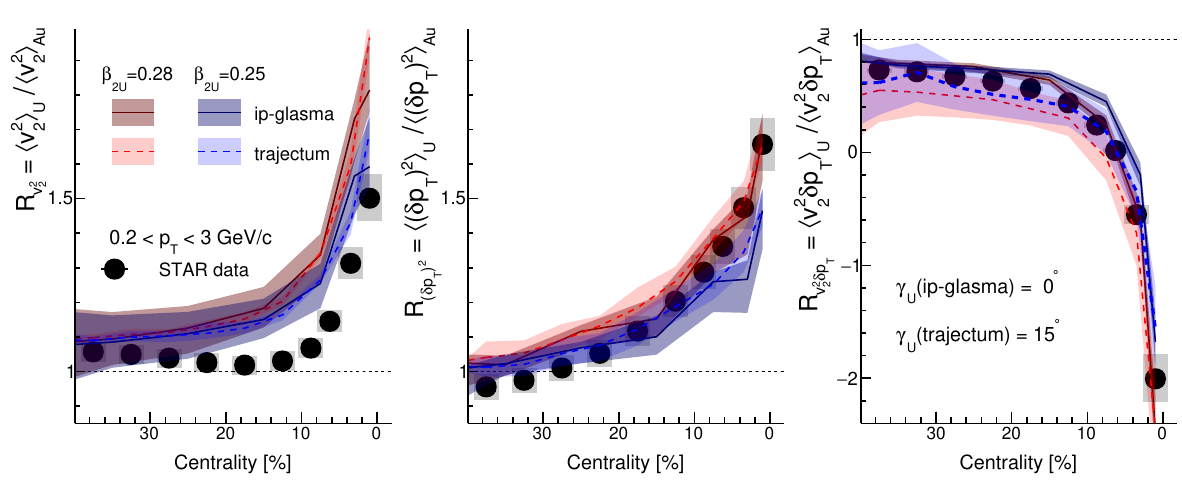}
\end{center}
\caption{\label{fig:a6} \textbf{Comparison between two hydrodynamic models}. The ratios of $\lr{v_2^2}$ (left), $\lr{(\delta\pT)^2}$ (middle) and $\lr{v_2^2\delta\pT}$ (right) as a function of centrality from IP-Glasma+MUSIC (solid lines) and Trajectum (dashed lines), assuming $\beta_{2\rm U}=0.28$ (red) and $\beta_{2\rm U}=0.25$ (blue). For the IP-Glasma+MUSIC model, only the uncertainties for the default configuration are shown for clarity. They are compared with the data.}
\end{figure*}

Model uncertainties for the ratios are derived by combining the impact of varying viscosities, together with various sources from Fig.~\ref{fig:a5}. As a consequence, checks that are consistent with the default calculation within their statistical uncertainties do not contribute to the model uncertainties. The combined model uncertainties for one standard deviation are displayed in Fig.~\ref{fig:3}.

A cross-check is conducted for an alternative hydrodynamic code, the Trajectum model~\cite{Nijs:2023yab,Giacalone:2023cet}. This model has 20 parameter sets obtained from a Bayesian analysis of the Pb+Pb data at the LHC, but was not tuned to the RHIC data.  For this calculation, we simply repeat the calculation at RHIC energy and calculate the same observables. While the description of the $\lr{v_2^2}$ and $\lr{(\delta \pT)^2}$ is reasonable, several parameter sets give negative values of $\lr{v_2^2\delta \pT}$ in mid-central collisions, and subsequently not used. The calculation is performed for the remaining 16 parameter sets as a function of centrality, and root mean square variations among these calculations are assigned as the uncertainty. 

Figure~\ref{fig:a6} displays the ratios of flow observables from Trajectum and compares them to IP-Glasma+MUSIC. The results from these two models agree within their uncertainties for $R_{v_2^2}$ and $R_{(\delta \pT)^2}$, with Trajectum predictions slightly higher in the UCC region. This leads to slightly lower values of $\beta_{2\mathrm {U}}$ than the IP-Glasma model: $\beta_{2\mathrm{U}}=0.228\pm0.013$ for $R_{v_2^2}$ and $\beta_{2\mathrm{U}}=0.276\pm0.018$ for $R_{(\delta \pT)^2}$.

For $R_{v_2^2\delta \pT}$, however, the Trajectum model tends to systematically under-predict the data, as well as has much larger uncertainties compared with IP-Glasma model. In central collisions, such discrepancy can be improved by using a larger triaxiality parameter value $\gamma_{\mathrm{U}}\sim 15^{\circ}$. Overall, the comparison of the Trajectum model with data gives similar constraints on $\beta_{2\mathrm {U}}$ with comparable uncertainties but larger $\gamma_{\mathrm{U}}$ value with bigger uncertainties (next section).
\begin{table}[h]
\centering
\caption{\label{tab:1}  \textbf{Choices of Woods-Saxon parameters} in Eqs.~\eqref{eq:3} and \eqref{eq:a9}, including deformations, in the IP-Glasma+MUSIC model. The default values are denoted by bold font, and the rest are variations designed to constrain the values of $(\beta_{2\mathrm{U}},\gamma_{\mathrm U})$ and derive theoretical uncertainties associated with other structure parameters.\\}
\hspace*{-0.3cm}\footnotesize{
\begin{tabular}{c|c|c|c|c|c|c}\hline 
Species                   & R(fm)              & a (fm)      & $d_{\mathrm {min}}$(fm) & $\beta_2$          & $\beta_4$      & $\gamma$ ($^{\circ}$) \\\hline
$^{197}$Au                 & {\bf 6.62},6.38   & {\bf 0.52}               &{\bf 0.9}, 0.4                  & {\bf 0.14},0.12   & {\bf 0}        & 53, {\bf 45}, 37 \\\cline{1-7}
\multirow{3}{*}{$^{238}$U} & \multirow{3}{*}{{\bf 6.81},7.07} & {\bf 0.55} &\multirow{3}{*}{{\bf 0.9}, 0.4} & 0,0.15           & \multirow{3}{*}{{\bf 0.09},0}   & {\bf 0}, 10 \\
                          &                                  & 0.495            &                          & 0.22, 0.25       &               &  15, 20 \\
                          &                                  & 0.605            &                          & {\bf 0.28},0.34  &               &   \\\hline
\end{tabular}}\normalsize
\end{table}

\subsection{Assigning uncertainties on $\beta_{2\mathrm U}$ and $\gamma_{\mathrm U}$}
A standard pseudo-experiment procedure, similar to that described in Ref.~\cite{Nisius:2014wua}, is employed to combine the uncertainties from $R_{(\delta \pT)^2}$ and $R_{v_2^2\delta \pT}$ shown in Fig.~\ref{fig:3}g. We assume that the total uncertainties extracted from the two observables are independent, and we model the probability density function as follows:
\begin{align}\label{eq:a8}
P(\beta_{2\mathrm U},\gamma_{\mathrm U}) \propto \exp\left(-\frac{(\beta_{2\mathrm U}-\bar{\beta}_a)^2}{2 \sigma_a^2}-\frac{(\beta_{2\mathrm U}-\bar{\beta}_b(\gamma_{\mathrm U}))^2}{2 \sigma_b^2(\gamma_{\mathrm U})}\right)\;.
\end{align}
Here, $\bar{\beta}_a=0.294$ and $\sigma_a=0.021$ represent the mean and uncertainty of $\beta_{2\mathrm U}$ extracted from $R_{(\delta \pT)^2}$ in Fig.~\ref{fig:3}g from the IP-Glasma+MUSIC model. Similarly, $\bar{\beta}_b$ and $\sigma_b$ are the mean and uncertainty of $\beta_{2\mathrm U}$ from $R_{v_2^2\delta \pT}$, and they depend on the parameter $\gamma_{\mathrm U}$. We sample a uniform prior distribution in $\beta_{2\mathrm U}$ and $\gamma_{\mathrm U}$ to obtain the posterior distribution. From this posterior distribution, we obtained the mean and one standard deviation uncertainty of $\beta_{2\mathrm U}$ and $\gamma_{\mathrm U}$, $\beta_{2\mathrm{U}}=0.297\pm0.015$ and $\gamma_{\mathrm{U}}=8.5^{\circ}\pm4.8^{\circ}$, as well as the confidence contours displayed in Fig.~\ref{fig:3}g. This statistical analysis is also performed for $R_{(\delta \pT)^2}$ and $R_{v_2^2\delta \pT}$ for the Trajectum model, yielding constraints of $\beta_{2\mathrm{U}}=0.275 \pm 0.017$ and $\gamma_{\mathrm{U}}=15.5^{\circ} \pm 7.8^{\circ}$. 

Finally, we perform an analysis to combine the constraints of the IP-Glasma+MUSIC and Trajectum models. This is achieved by multiplying the probability density function Eq.~\eqref{eq:a8} from the two models, treating their constraints as statistically independent.  This approach yields $\beta_{2\mathrm{U}}=0.286 \pm 0.012$ and $\gamma_{\mathrm{U}}=8.7^{\circ} \pm 4.5^{\circ}$. We noticed that Trajectum model does not impact the constraints on $\gamma_{\mathrm{U}}$ due to the model's large uncertainty, but the uncertainty on $\beta_{2\mathrm{U}}$ reduces significantly due to the comparable precision in the two models. Therefore, we also include the difference of the extracted $\beta_{2\mathrm{U}}$ values between the two models as an additional theoretical uncertainty. The final constraints given by this procedure are $\beta_{2\mathrm{U}}=0.286 \pm 0.025$ and $\gamma_{\mathrm{U}}=8.7^{\circ} \pm 4.5^{\circ}$.

\subsection{Data availability}
All raw data for this study were collected using the STAR detector at Brookhaven National Laboratory and are not available to the public. Derived data supporting the findings of this study are publicly available in the HEPData repository (https://www.hepdata.net/record/147196) or from the corresponding author on request.

\subsection{Code availability}
The codes to process raw data collected by the STAR detector are publicly available on GitHub (https://github.com/star-bnl). The codes to analyse the produced data are not publicly available.

\bibliography{ref}{}

\subsection{Acknowledgments}
We thank the RHIC Operations Group and RCF at BNL, the NERSC Center at LBNL, and the Open Science Grid consortium for providing resources and support.  This work was supported in part by the Office of Nuclear Physics within the U.S. DOE Office of Science, the U.S. National Science Foundation, National Natural Science Foundation of China, Chinese Academy of Science, the Ministry of Science and Technology of China and the Chinese Ministry of Education, the Higher Education Sprout Project by Ministry of Education at NCKU, the National Research Foundation of Korea, Czech Science Foundation and Ministry of Education, Youth and Sports of the Czech Republic, Hungarian National Research, Development and Innovation Office, New National Excellency Programme of the Hungarian Ministry of Human Capacities, Department of Atomic Energy and Department of Science and Technology of the Government of India, the National Science Centre and WUT ID-UB of Poland, the Ministry of Science, Education and Sports of the Republic of Croatia, German Bundesministerium f\"ur Bildung, Wissenschaft, Forschung and Technologie (BMBF), Helmholtz Association, Ministry of Education, Culture, Sports, Science, and Technology (MEXT), Japan Society for the Promotion of Science (JSPS) and Agencia Nacional de Investigaci\'on y Desarrollo (ANID) of Chile. We thank C. Shen for providing the IP-Glasma+MUSIC code and G. Nijs for providing the Trajectum code. We thank G. Giacalone, D. Lee, T. Rodriguez, B. Schenke, H. Song, and Y. Zhou for their discussions and valuable comments. 


\end{document}